\documentstyle[11pt,newpasp,twoside,epsf]{article}
\def\edcomment#1{\iffalse\marginpar{\raggedright\sl#1\/}\else\relax\fi}
\reversemarginpar

\begin{document}
\title{X--ray bright optically quiet galaxies: the case of P3}
 \author{Andrea Comastri, Marcella Brusa, Paolo Ciliegi, Marco Mignoli, Cristian Vignali}
\affil{Osservatorio Astronomico di Bologna, via Ranzani 1, 40127 
Bologna, Italy}
\author{Paola Severgnini, Roberto Maiolino}
\affil{Dipartimento di Astronomia Universita' di Firenze, Italy}
\author{Fabrizio Fiore}
\affil{Osservatorio Astronomico di Roma, Italy}
\author{Fabio La Franca, Giorgio Matt, G. Cesare Perola}
\affil{Universita' Roma Tre, Roma, Italy}
\author{Alessandro Baldi, Silvano Molendi}
\affil{IFCTR/CNR, Milano, Italy}

\begin{abstract}
Recent X--ray surveys have clearly demonstrated that a population 
of optically dull, X--ray  bright galaxies is emerging at X--ray 
fluxes of the order of $10^{-14}$ erg cm$^{-2}$ s$^{-1}$.
The nature of these objects is still unknown.
We present the results of multiwavelength  
observations of what can be considered the best studied example:
the {\it Chandra} source CXOUJ031238.9--765134
optically identified by Fiore et al. (2000) with an apparently normal
galaxy at $z$=0.159 and called ``{\tt FIORE P3}''.                 
\end{abstract}

\section{Introduction}

The unprecedent arcsec {\it Chandra} spatial resolution
allowed to unambiguosly identify a class of X--ray emitting
sources associated with the nuclei of optically
``normal'' galaxies without any obvious signature of AGN activity
(Mushotzky et al. 2000, Barger et al. 2001, Fiore et al. 2000 
(hereinafter F00), Giacconi et al. 2001, Hornschmeier et al. 2001).  
The high X--ray--to--optical flux ratio and the
hard X--ray spectra both suggest the presence of an
obscured AGN. Unfortunately, most of the sources are detected
with a number of photons which is too low
to apply conventional X--ray spectral fitting techniques and
to constrain the absorbing column density.
As a consequence, alternative possibilities
(besides an obscured AGN) are still viable.

In this paper we present a multiwavelength study of
the X--ray source CXOUJ 031238.9--765134 (hereinafter P3, being the 
third source catalogued by F00 in the field containing 
the quasar PKS 0312--76)  
optically identified as a normal early type galaxy at $z$=0.159.
 
The observations have been carried out as a part of a program
aimed to understand the nature of the sources responsible for most
of the hard X--ray background by means of multiwavelength
follow--up observations of hard X--ray selected sources
serendipitously discovered in the XMM--{\it Newton} field of view.
The survey (named {\tt HELLAS2XMM} as extend the
BeppoSAX High Energy Large Area Survey to XMM--{\it Newton})
is described elsewhere in this volume 
(Fiore et al. and Baldi et al.). 
Here we briefly report on the multiwavelength observations 
of P3 carried out in the X--ray, optical, infrared and radio 
bands and present the overall broad band energy distribution. 
A more detailed description of the observations and the results 
is reported in Comastri et al. (2001).

\section{Multiwavelength observations}

\begin{figure}
\plottwo{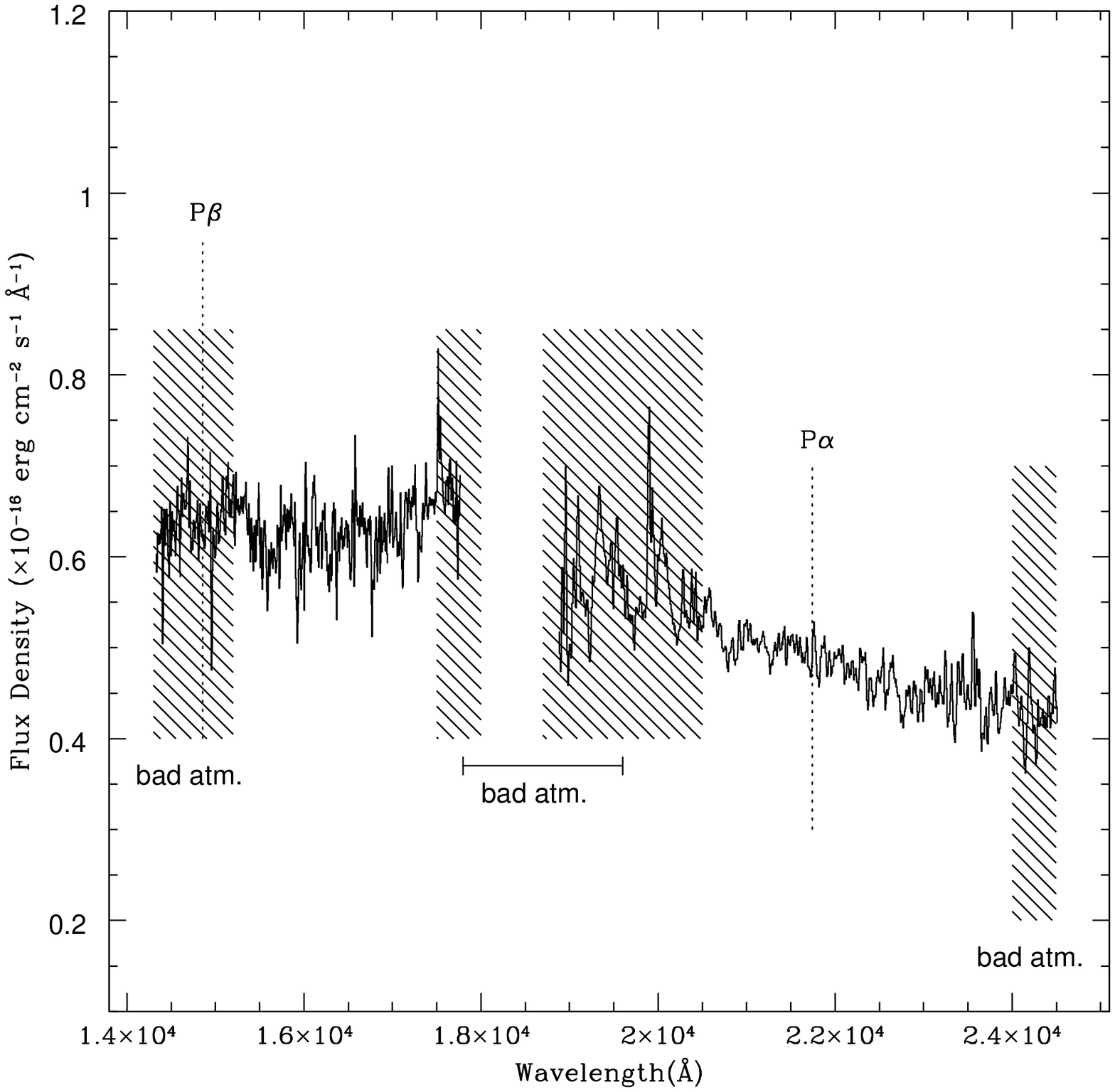}{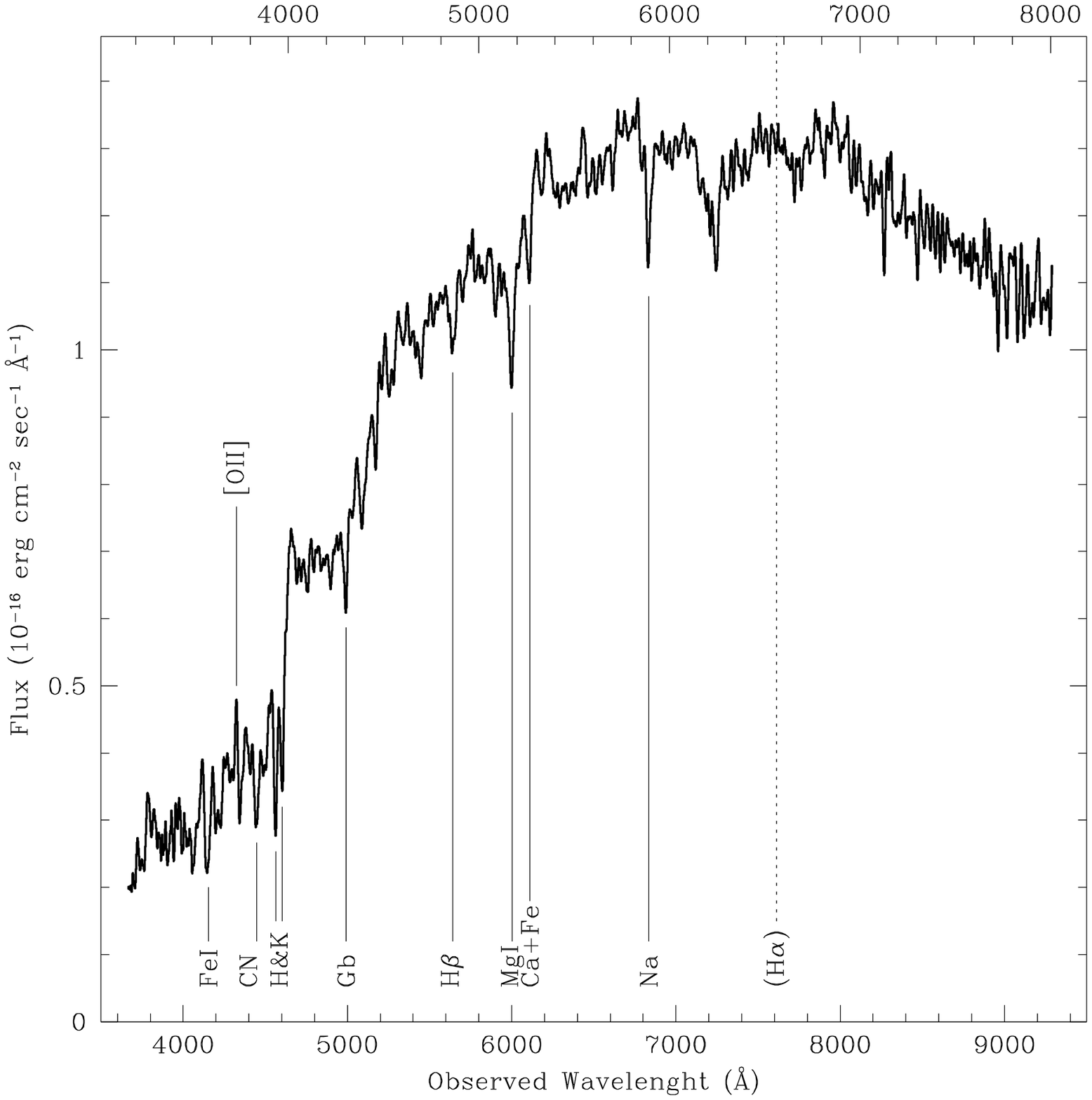}
\caption{The flux calibrated near--infrared (left panel) and optical
spectra (right panel). The shaded regions correspond to wavelengths of
bad atmospheric transmissions. The expected positions of H$\alpha$, 
P$\alpha$ and P$\beta$ are also indicated.}
\end{figure}

Radio observations were performed at 5 GHz with the Australia Telescope
Compact Array (ATCA). The data were analysed with the software 
package {\tt MIRIAD}. We searched for a possible 
radio counterpart of P3 within a box of 20 $\times$ 20 arcsec centered on
the X--ray position determined by {\it Chandra} where the noise 
is relatively uniforme at an average level of 50 $\mu$Jy. 
No sources were found and thus the 3$\sigma$ upper limit on the radio 
flux is 0.15 mJy.
  
Near Infrared spectroscopic observations 
were performed at the ESO Very Large Telescope (VLT).
The galaxy has been observed with the 1\arcsec ~ slit in two
different different filters, SH (1.42-1.83 $\mu$m) and SK (1.84-2.56 $\mu$m),
available for the Low Resolution grating (LR) in the Short Wavelength (SW)
configuration of ISAAC.  
The corresponding H and Ks magnitudes are 15.7$^m$ and 14.9$^m$ respectively. 
The final spectra in both the filters 
are shown in the left panel of Fig.~1 
(precision on the flux calibration $\sim$ 10\%).
Additional imaging observations in the L band have been performed
with the ISAAC camera. The source is not detected at the limiting
magnitude of 13.4$^m$ (3$\sigma$ upper limit).

\begin{figure}
\plotone{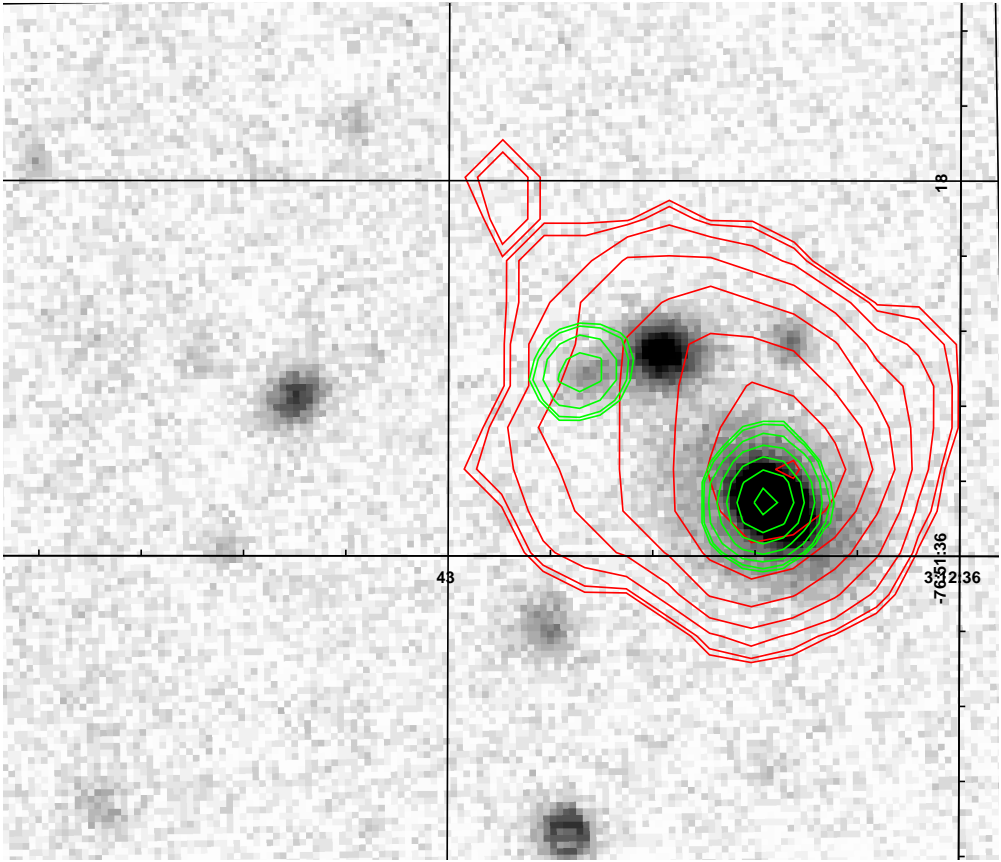}
\caption{The {\it Chandra} (green/light grey) and XMM--{\it Newton}
(red/black) contours overlayed on the EFOSC R band image}
\end{figure}

Optical spectroscopy was performed with the ESO 3.6m
telescope equipped with EFOSC2 during two different observing runs.
The comparison of spectra taken one year apart shows a reasonable good
agreement, compatible with the precision of the flux calibration ($\sim$ 
15\%). In the righr panel of Fig.~2 we show the average of the two spectra, 
with the 
principal identified absorption lines labelled. Also plotted are the
emission line [O{\tt II}] at 3727 \AA\ and the expected position of the
H$\alpha$ features. Unfortunately, at the redshift of P3 the H$\alpha$ line
position coincides with the stronger atmospheric telluric band in the optical.
The magnitudes in the R and B bands are 18.0$^m$ and 19.7$^m$
respectively. The average redshift is $z$ = 0.1595$\pm$0.0007, 
measured from the position of the principal absorption lines and 
comparison with an early--galaxy template.

The {\it Chandra} observation reported by F00 has been combined with an
additional exposure of the same field and analysed using version 2.1
of the CXC software.
The 2--10 keV flux, computed from the exposure corrected 
image assuming a power law spectrum of $\Gamma$= 1.8 and Galactic absorption,
(8 $\times$ 10$^{20}$ cm$^{-2}$) 
is 2.6 $\times$ 10$^{-14}$ erg cm$^{-2}$ s$^{-1}$.
Although the counting statistic is too poor to allow spectral fitting
the number of counts detected in the 2--8 keV and 0.5--2 keV bands 
suggest a hard spectrum.
Our target was clearly detected also in the 25 ksec XMM--{\it Newton}
PV observation of the same field.
Even though the XMM positional accuracy is not as good as the
{\it Chandra} one, the two X--ray centroids agree within about 2 arcsec.
The XMM and {\it Chandra} contours are overlaid on a R band optical image
in Fig.~2. It can be seen that the elongated XMM contours are probably
due to the presence of a faint X--ray source which is clearly
resolved by {\it Chandra} at a distance of $\sim$ 6 arcsec from P3. 

\begin{figure}
\plotone{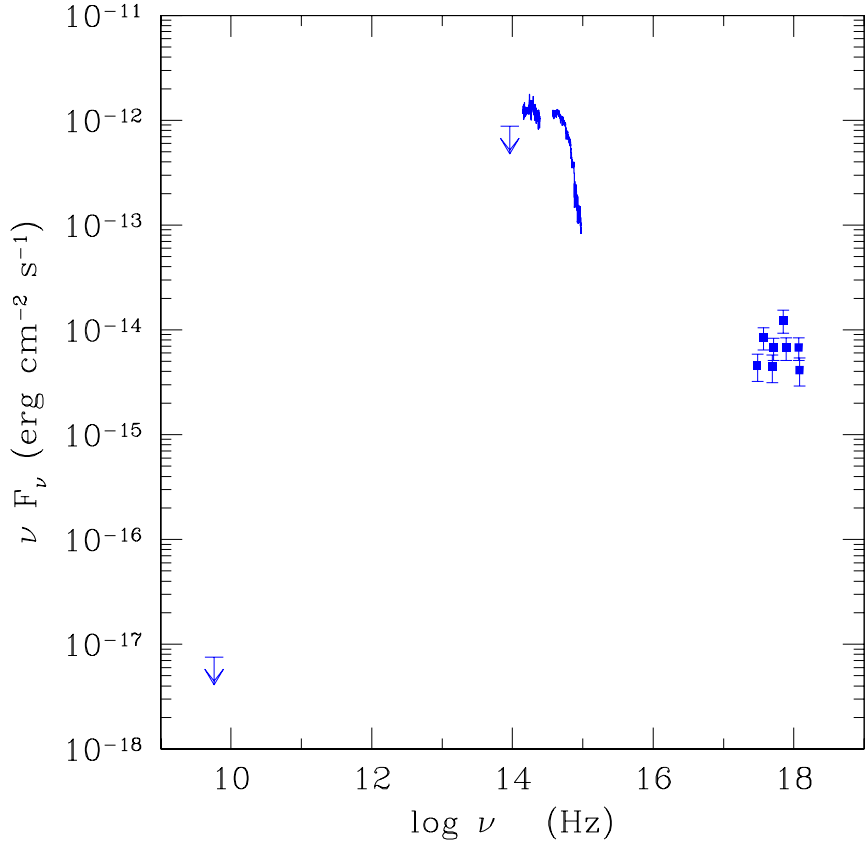}
\caption{The radio to X--ray spectral energy distribution.}
\end{figure}

The XMM--{\it Newton} data were analyzed 
using the latest version of the SAS software. 
The PN and MOS spectra were extracted from the cleaned event files, 
(see Baldi et al. 2001 for a description of data reduction)
 rebinned with at least 20 counts per channel
and fitted simultaneously using XSPEC leaving the relative normalizations
free to vary. The latest version of response and effective area files
was used. The spectrum is hard also in the XMM--{\it Newton} 
observation being equally well
fitted by a flat power law with photon index $\Gamma$ = 1.10$\pm$0.35 
or by a steeper spectrum with the slope fixed at $\Gamma$=1.8 
(a typical value for AGN) plus an intrinsic column density 
$N_H$ = 8$\pm$5 $\times$ 10$^{21}$ cm$^{-2}$ at the source frame.
The de--absorbed X--ray luminosity in the hard 2--10 keV  band is about 
3 $\times$ 10$^{42}$ erg s$^{-1}$.

\section{What's going on in the nucleus of P3 ?}

The broad band observations discussed above are shown in 
a $\nu$ vs $\nu F_{\nu}$ plot in Fig.~3.
At the first sight the overall spectral energy distribution 
(hereinafter SED) is clearly dominated by the optical--infrared light
of the host galaxy.
The X--ray flux level is however almost two order of magnitude greater 
than that expected on the basis of the $L_X$--$L_B$ correlation 
of early type galaxies (Fabbiano et al. 1992). 
The relatively high X--ray luminosity and the X--ray spectral properties
strongly suggest nuclear activity in the central regions of P3.
Based on the {\it Chandra} detection and the optical spectrum F00
suggest three different possibilities namely: 1) a radiatively inefficient 
advection dominated accretion flow (ADAF), 2) a BL LAC object, 
3) a completely obscured AGN. 
The multiwavelength observations presented in this paper allow us to 
investigate in more detail these possibilities.

Although the hard X--ray spectrum is consistent with the flat 
power law slope expected by (some) ADAF models (see Di Matteo et al. 2000), 
the model predicted radio flux is in excess of the present upper limit.

The BL Lac hypothesis might be tenable 
despite the presence of a large Calcium break and 
the low radio flux density. 
Indeed there is increasing evidence that the Blazars SED 
can be unified in a spectral sequence determined from the total luminosity 
(Fossati et al. 1998, Ghisellini et al. 1998). The objects populating
the low--luminosity extreme of the sequence are called HBL 
(High peak BL Lacs) as their SED peak at relatively high frequency
(typically in the X--rays) when compared to that of e SED of other blazars.
The upper limit on the 5 GHz radio luminosity density ($<$ 10$^{39}$ 
erg s$^{-1}$), on the radio to X--ray spectral index ($\alpha_{rx} < 0.6$)
and the flat slope in the X--ray band  
would be consistent with the presence of a rather extreme member of the 
HBL class hosted by the P3 galaxy.
    
\begin{figure}
\plottwo{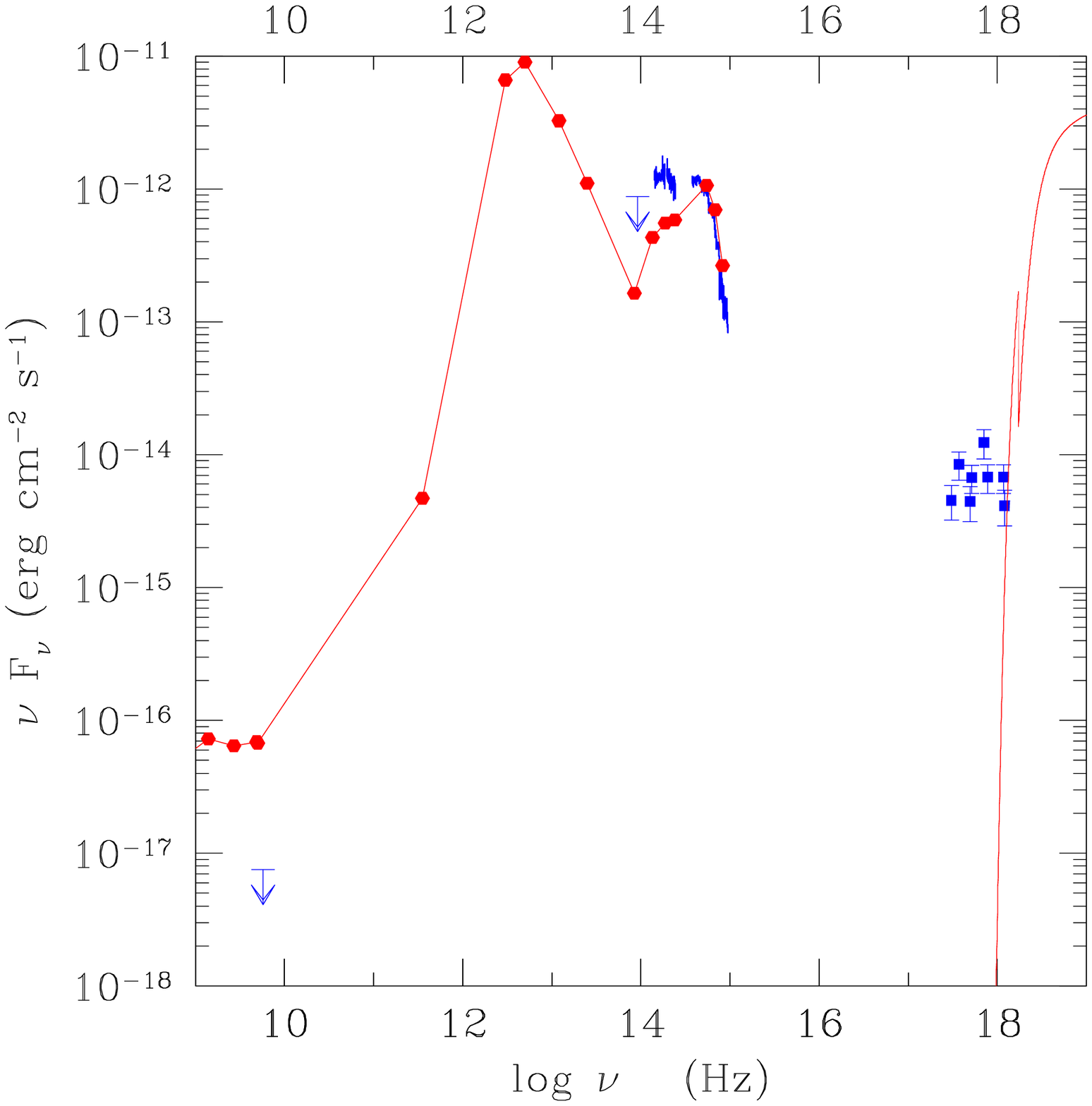}{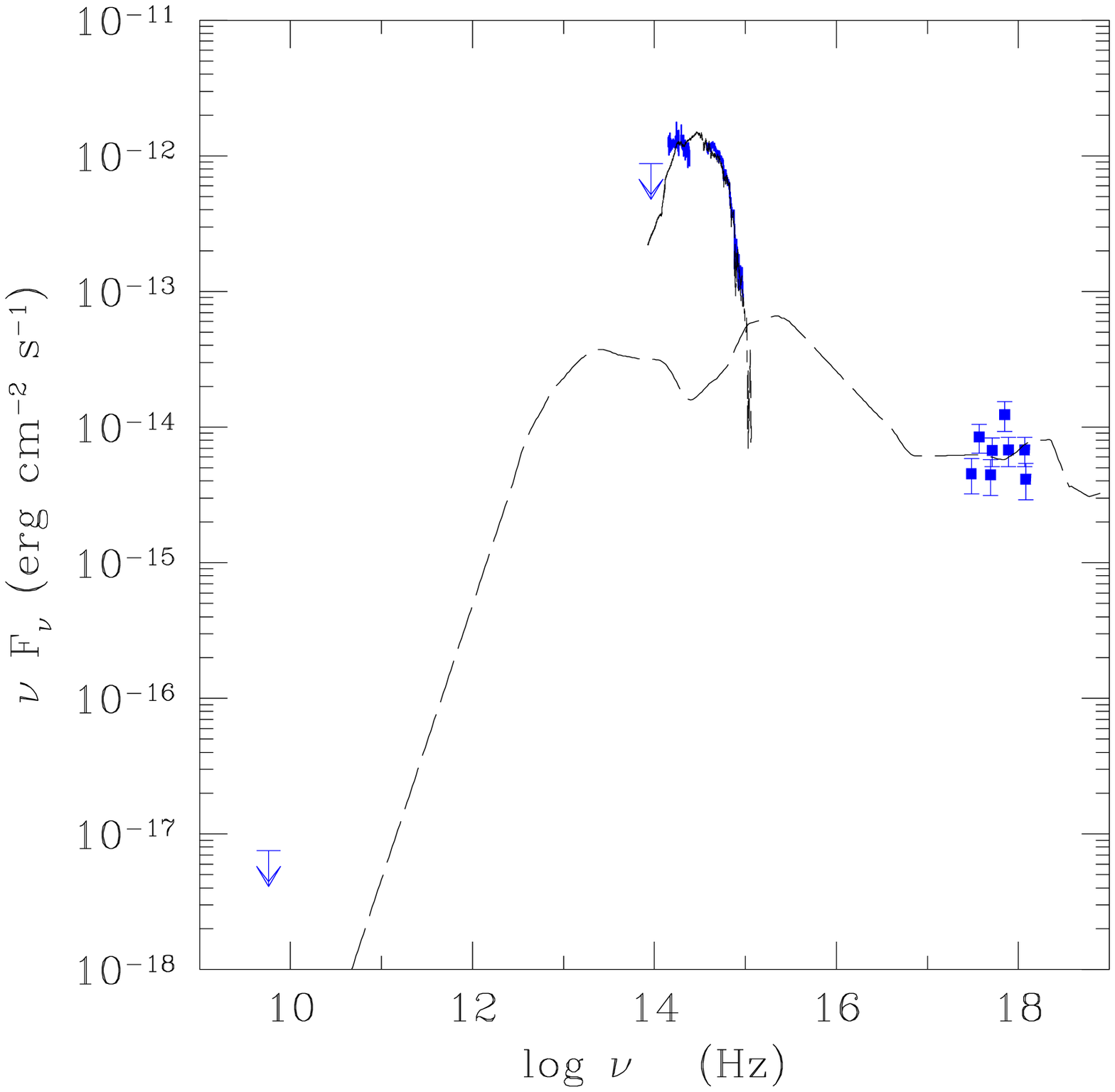}
\caption{The P3 SED compared with that of the highly obscured 
Seyfert 2 galaxy NGC 6240 (left) and with the average SED of a 
radio quiet AGN plus an early galaxy template (right)}
\end{figure}

An alternative explanation would be the presence of an hidden 
AGN. There are several examples of (obscured) AGN discovered only 
by means of X--ray observations and not recognized by optical spectroscopy
(see Matt, this volume, and reference therein). In order to test whether this 
is the case for P3 we have compared the observed SED (Fig.~4, left panel) 
with that of NGC 6240: a highly obscured AGN.
The AGN SED is normalized to match the P3 optical flux.
Besides the disagreement at long wavelengths which can be due to
a more intense star--formation in NGC 6240, the P3 multiwavelength data 
are in relatively good agreement with the obscured AGN template.
If P3 hosts a Compton thick AGN with a SED similar to that of NGC 6240
then a luminous X--ray source should be present at higher energies.
The observed SED would be also consistent with an early type galaxy template
normalized to match the optical and infrared spectrum plus the average 
SED of radio--quiet AGN (Elvis et al. 1994) rescaled to the observed
X--ray flux (Fig.~4, right panel). In this case the optical AGN emission lines
are diluted by the host galaxy starlight and could be revealed by 
narrow--slit optical spectroscopy with HST.  

In both the cases further X--ray and optical observations are required to 
understand the nature of P3 and of the X--ray bright optically quiet
galaxy population.


\begin{references}
\reference Baldi A., Molendi S., Comastri A., Fiore F., Matt G., Vignali C., 
 2001, \apj, in press (astro--ph/0108514)
\reference Barger A., Cowie L., Mushotzky R.F., Richards E.A., 2001, \aj, 121, 662
\reference Comastri A., et al. 2001, in preparation 
\reference Di Matteo T., Quataert E., Allen S.W., Narayan R., Fabian A.C., 
 2000, \mnras, 311, 507  
\reference Elvis M., Wilkes B.J., McDowell J.C., et al. 1994, \apjs, 95, 1
\reference Fabbiano G., Kim D.W., Trinchieri G., 1992, \apjs, 80, 531
\reference Fiore F., La Franca F., Vignali C., et al. 2000, (F00) New 
Astronomy 5, 143
\reference Fossati G., Maraschi L., Celotti A., Comastri A., Ghisellini G.,
1998, \mnras, 299, 433  
\reference Ghisellini G., Celotti A., Fossati G., Maraschi L., Comastri A., 
1998, \mnras, 301, 451 
\reference Giacconi R., Rosati P., Tozzi P., et al. 2001, \apj, 551, 664  
\reference Hornschemeier A., et al. 2001, in press (astro--ph/0108228)
\reference Mushotzky R.F., Cowie L.L., Barger A.J., Arnaud K.A., 2000, Nature, 404, 459 
\end{references}
\end{document}